\documentclass[aps,prb,twocolumn,floats]{revtex4-2}
\usepackage{epsfig}
\usepackage{amsmath}
\usepackage{amssymb}
\usepackage{graphicx}
\usepackage{soul}
\usepackage{color}

\begin{document}
\title{Transport through a Nodal Surface Semimetal - Superconductor junction in absence/presence of light irradiation}
\author{Bhaskar Pandit$^1$}
\email{bhaskar.pandit1994@gmail.com}
\author{Satyaki Kar$^2$}
\email{satyaki.phys@gmail.com  (corresponding author)}
\affiliation{$^1$Netaji Mahavidyalaya, Arambagh, West Bengal - 712601, India\\$^2$AKPC Mahavidyalaya, Bengai, West Bengal -712611, India}
\begin{abstract}
  We study quantum tunnelling via $s$-wave superconductor (SC) junction with a topologically charged nodal surface semimetal (NSSM) where a nonsymmorphic symmetry forces the nodal surfaces to stick to the Brillouin Zone boundary. Due to their unique dispersions close to the two dimensional band crossing, the charge carriers in the NSSM display many unorthodox behavior in the nature of Andreev as well as normal reflections at the SC junction interface. We investigate such behaviors for different incident orientations for both subgap and supergap energies where monotonic decays/rises of reflectance with incident energy or angle of incidences are often not followed. We also consider irradiation via light with circular and linear polarization on such systems following a Floquet approach in the limit of high frequency irradiation and probe the stroboscopic temporal evolution of the transport parameters. Our results indicate many nontrivial Andreev transport features including near-depletion of the subgap conductivities.
All these nontrivialities can be tested in a cold atom set-up on optical lattices and well experimented for quantum information processing purpose.
\end{abstract}
\maketitle                              
\section{Introduction}

Topological robustness\cite{wen} appears to be the holy grail for the condensed matter physics community as long as the nectar of disorder-independent charge transport is in demand. That motivates both the physicists and engineers to be in constant look out for different means or nature of nontrivial band crossings or avoided crossings in various materials and {numerous attempts have been made to discover potential material candidates that exhibit such exotic phenomena}. Chronologically, such crossings were observed first in {zero-dimensional} ($e.g.,$ Graphene, Weyl semimetals)\cite{armitage,skrev} and {one-dimensional} ($e.g.,$ nodal line semimetals)\cite{fang} systems within the momentum space, but 2D band crossings have also been reported lately encouraging thorough investigation of nodal surfaces (NS) and nodal surface semimetals (NSSM).

A two dimensional (2D) crossing of bands with linear dispersion away from it constitutes what we call a nodal surface\cite{wu,zhong,turker}. They often bear the signature of symmetry protected topological charges amounting to nontrivial phases\cite{yang,xiao,jpcm}.
In this regard, a non-symmorphic symmetry\cite{furusaki,liang} (a combination of a point group symmetry and half-lattice translation) is very relevant as this, in combination with a time reversal symmetry $\it{T}$ can lead to a two fold degeneracy at the 2D boundary (say, $k_z=\pi$) of the 3D Brillouin zone\cite{liang,wu}. These band crossings are topologically robust both locally and globally\cite{zhao}.

{So far, the discussion on such nodal-surface semimetals (NSSMs) have
found place only in a few isolated theoretical studies: A
family of stable Graphene network materials, each displaying a pair of nodal surfaces close to the Fermi level,
was proposed by Zhong $et.~al.$\cite{zhong}. Liang $et.~al.$\cite{liang} found that
hexagonal $\it{BaMX_3}~ (M = V, Nb,~or~ Ta;~ X = S~ or~ Se)$
exhibits a single nodal surface when spin-orbit coupling
is not considered. In a proposed acoustic metamaterial\cite{xiao}, a nodal surface similar to that of Ref.\cite{liang} was predicted, and its stability under various perturbations
was later analyzed in a theoretical study\cite{turker}.
}

This paper {is devoted to the study of} the junctions of superconductors with such exotic NSSM systems and quantum transport through it.
Generally charge transport through a $s$-wave superconductor - normal metal (SN) junction is characterized by Andreev reflections\cite{andreev} (AR), where a hole is reflected back from the interface towards the normal metal side predominantly for subgap incident energies: $E<\Delta$ with $\Delta$ being the superconducting pair potential. Though an usual intraband electron-hole conversion leads to retro AR (RAR), Dirac-like spectra {with unconventional Fermi surfaces} adds {significant variations} to the tunnelling transport phenomena. Interband specular AR (SAR) are predicted at low energies in a Graphene based SN junction\cite{beenakker} and then in other junctions involving Dirac materials like Silicene\cite{silicene}, $MoS_2$\cite{mos2} or Phosphorene\cite{phosphorene} etc. In this respect, superconducting junctions with topological semimetals are also very relevant. Relative orientation of point nodes in Weyl semimetals (WSM) with respect to the incident stream of carriers produces anisotropic Andreev conductance in a $s$ wave superconductor junction\cite{wsm} while in a nodal line semimetal (NLSM), even double Andreev reflections can be observed\cite{prb101}. The transport features of superconducting junctions with a NSSM is rather not investigated as yet. {However, due to the 2D nodal surfaces and subsequent unique low energy dispersions around nodal surfaces, one can naturally expect many nontrivial transport behaviors for quantum tunnelling through such junctions. Hence} we attempt to probe the same in this work.
Like in a WSM or a NLSM, here also different relative orientations of the nodal surface with respect to the interface with superconductor give rise to distinct Andreev transport features. However in this work, we have only restricted the study to NS perpendicular to the junction interface.

A light irradiation to NSSM brings in {time periodic fields to the problem and the dynamics of the charge carriers get modulated accordingly.} Observing stroboscopically, $i.e.,$ in steps of periodicity of the fields in the irradiation, an effective stationary Floquet Hamiltonian can be constructed\cite{eckardt,moessner} depicting the dynamics under periodic variation of the fields. Such a Floquet expansion can unravel interesting dynamic feature depending on the polarization direction of the radiation\cite{debu} and thus such irradiation on a superconducting junction involving a NSSM, results in interesting modification in transmitivity and reflectivities as can be examined in the Floquet space.

In this report we discuss the quantum transport across a superconductor junction with a topologically charged NSSM in Section II. Later in Section III, we consider light irradiation on such system and investigate their dynamic behavior using a Floquet-Magnus analysis\cite{eckardt} for a circular (Section IIIA) and linear (Section IIIB) polarization. Lastly in Section  IV, we summarize our results and brief on further scopes of our work.

\begin{figure}
   \vskip -.3 in
  \begin{center}
   \begin{picture}(100,100)
     \put(-80,0){
        \includegraphics[width=.34\linewidth, height= 1.2 in]{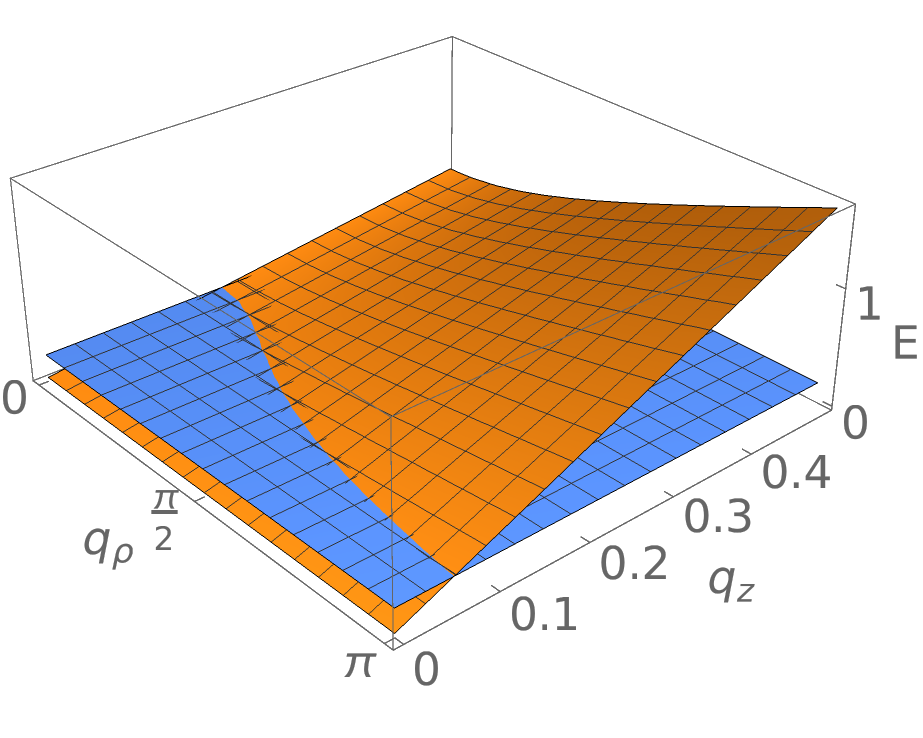}
        \includegraphics[width=.34\linewidth, height= 1.05 in]{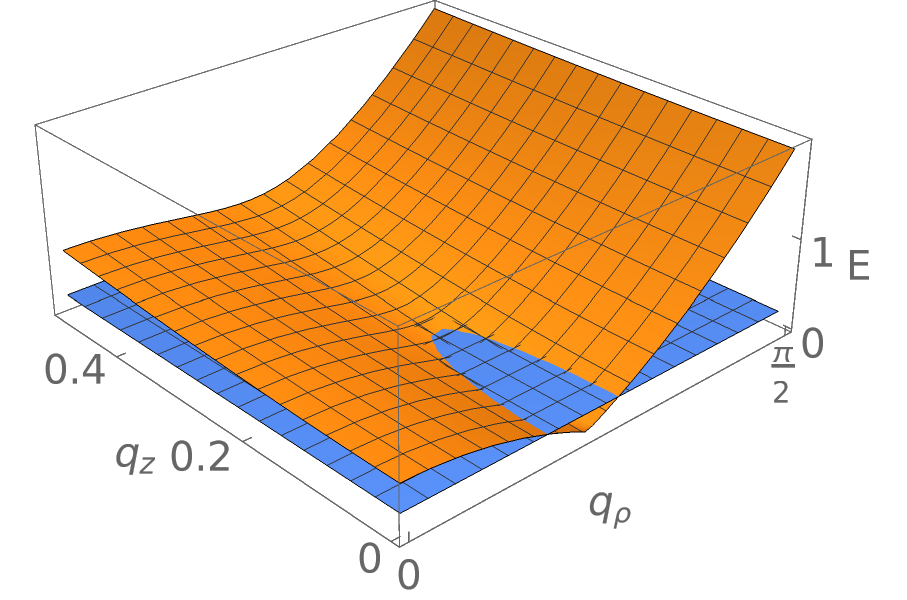}
        \includegraphics[width=.34\linewidth, height= 1.05 in]{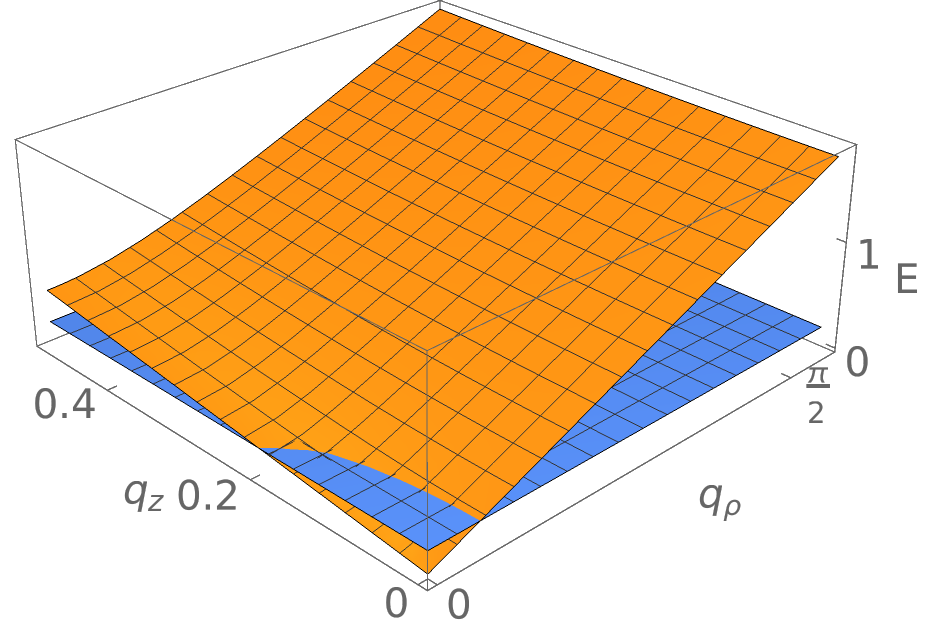}}
     \put(-65,60){(a)}
     \put(15,60){(b)}
     \put(105,60){(c)}
   \end{picture}  
  \end{center}
 \vskip -.2 in
 \caption{Typical cartoons for low energy dispersions around a (a) NSSM, (b) NLSM and a (c) WSM
   and their intersection with constant energy surface at $E=0.2$ in a $q_z$-$q_\rho~(=\sqrt{q_x^2+q_y^2})$ plane. Here $E=0$ corresponds to (a) a nodal surface at $q_z=0$, (b) a nodal circle at $q_z=0,q_\rho\ne0$ and (c) a Weyl point at $q_z=q_\rho=0$. At a low energy $E=0.2$, an electron at a constant nonzero $q_z$ say $q_z=0.1$, should carry a single $q_\rho\sim\pi/2$ in the NSSM or a single $q_\rho \sim0$ in the WSM. But in the NLSM, the electron can carry two different $q_\rho$ values supporting double reflections in a NLSM-SC junction\cite{prb101}.} 
\label{disp}
\end{figure}

\section{An NSSM-SC Junction}

As mentioned above, here we describe the scattering processes and conductance via a junction of a NSSM material with a proximity induced $s$-wave superconductor. We consider
the nodal surface to be perpendicular to the interface and the formalism is discussed accordingly. Particularly, we take the NSSM and SC to be located in the regions $x<0$ and $x > 0$, respectively, with the NSSM-SC interface being at $x = 0$. A step potential given as $V(x)=V_s\Theta(x)$ is considered that vanishes in the normal side.

In the NSSM considered, the NS is brought about by non-symmorphic symmetry where the NS appears usually at the BZ boundary. {In Ref.\cite{xiao}, and then in Ref.\cite{jpcm}, it has been shown how a tight-binding model can give rise to nodal surfaces due to non-symmorphic symmetry.} In the present case the NS is denoted by the $k_z=\pi$ plane\cite{xiao}. About a point $k_0=(0,0,\pi)$ on that plane, we can write the continuum model of the NSSM at $k=k_0+q$ as $H=H(q)=Aq_z(q_x\sigma_x+q_y\sigma_y)+Bq_z\sigma_z${\cite{xiao,jpcm}}. Here for simplicity we consider $A=B=1$ {resulting in
  \begin{equation}
    H=q_z(q_x\sigma_x+q_y\sigma_y)+q_z\sigma_z.
    \label{eq1}
    \end{equation}}
{By making a comparison with simple NLSM and WSM continuum models given by Hamiltonians $H_{nlsm}=(q_x^2+q_y^2-\Lambda)\sigma_x+q_z\sigma_z$ and $H_{wsm}=q_x\sigma_x+q_y\sigma_y+q_z\sigma_z$  respectively, one can see how the surface node appears in the NSSM at energy $E=0$, in contrast with the line node and point nodes appearing in the NLSM and WSM respectively (see Fig.\ref{disp}). Also note that at a small energy, say $E=0.2$, there is a single value of $q_\rho$ available for a small $q_z$ in a NSSM or a WSM. So for an electron incident at an interface with a SC at $x=0$, there will be a single equi-energy reflected electron with same $q_\rho$ value. Contrarily for a NLSM, two values of $q_\rho$'s are available for a constant small $E$ and $q_z$ and this leads to double reflections at the interface with a SC\cite{prb101}.}

So the Bogoliubov-de-Gennes (BdG) Hamiltonian of this {NSSM-SC junction} problem is given by
\begin{widetext}
\begin{displaymath}
\left(\begin{array}{cc}
    H+V(x)-\mu & \Delta(x)\\
    \Delta^*(x) & \mu-H-V(x)
  \end{array}\right)=
\left(\begin{array}{cccc}
 V(x)+q_z-\mu & q_xq_z-iq_yq_z & \Delta(x) & 0\\
    q_xq_z+iq_yq_z & V(x)-q_z-\mu & 0 & \Delta(x)\\
    \Delta^*(x) & 0 & \mu-q_z-V(x) & -(q_xq_z-iq_yq_z)\\
    0 & \Delta^*(x) & -(q_xq_z+iq_yq_z) & \mu+q_z-V(x)
    \label{bdg}
  \end{array}\right)
\end{displaymath}
\end{widetext}
where $\Delta(x)=\Delta\Theta(x)$ denotes the pairing potential of the superconductor.

 In the normal side ($x<0$), the quasiparticle dispersions of the NSSM for electrons/holes, that are obtained by solving the BdG equation, comes out to be
\begin{align}
E_e^\pm &=\pm q_z\sqrt{1+q_\rho^2}-\mu,~~
E_h^\pm &=\pm q_z\sqrt{1+q_\rho^2}+\mu
\end{align}
$\mu$ being the chemical potential and $q_\rho^2=q_x^2+q_y^2$.
 We consider electron transport along the $x$ direction.
Taking the conduction band $E_e^+$ as an example, when the incident energy $E$ and wave vectors $q_y$ and $q_z$ are given, the equation $E_e^+(q_x,q_y,q_z)=E$ of $q_x$ has only two solutions, $q_x^e$ and -$q_x^e$ . Here the $q_x^e$ and -$q_x^e$ states propagate along the $+x$ and $-x$ directions respectively and can be considered as incident and reflected wave pairs. So in this case, there is only a single normal reflection process (and also a single Andreev reflection coming from the solution of the hole band, unlike double reflections observed in NLSM-SC junctions\cite{prb101}). Consider that the incident electrons are from the conduction band $E_e^+$ . They will be specularly reflected as electrons on the band $E_e^+$ and retro-
Andreev reflected as holes on the conduction band $E_h^-$ if $E<\mu$ or specular Andreev reflected as holes on the valence band $E_e^+$ if $E>\mu$ . In this respect, the scattering processes are the same as that in the Graphene-SC junction \cite{beenakker, cheng, xing}.

So for an incident electron with the wave vector $q_x^e$ and energy $E$ obeying $\mu>E$, the wave function can be written as

\begin{align}
&\psi_N(x<0) = \psi_N^{e+} + r \psi_N^{e-} + r_A\psi_N^{h-}=\frac{ e^{iq_x^{e+}x}}{\sqrt{Re[\chi_{11}]}}\left(\begin{array}{c}
  1\\
  \chi_{11}\\
  0\\
  0\\
  \end{array}\right)\nonumber\\
  &  + r\frac{ e^{-iq_x^{e+}x}}{\sqrt{Re[\chi_{12}]}}\left(\begin{array}{c}
  1\\
  -\chi_{12}\\
  0\\
  0\\
  \end{array}\right) 
  + r_A \frac{ e^{-iq_x^{h-}x}}{\sqrt{Re[\chi_{22}]}}\left(\begin{array}{c}
  0\\
  0\\
  1\\
  \chi_{22}\\
  \end{array}\right)
  \label{wvfn1}
\end{align}

Where
\begin{align}\label{eq4}
\chi_{11}&=\chi_{11}^0e^{i\theta_e}=\sqrt{\frac{(E +\mu)-q_z}{(E +\mu)+q_z}} e^{i\theta_e},~
\chi_{12}=\chi_{11} e^{-2i\theta_e},\nonumber\\
\chi_{22}&=\chi_{22}^0e^{i\theta_A}=\sqrt{\frac{(E -\mu)+q_z}{(E -\mu)-q_z}} e^{i\theta_A}.
\end{align}
$\theta_e$ is the angle of electron incidence in the $xy$ plane with $r$ and $r_A$ being the normal and Andreev reflection coefficients respectively.
The denominators in the three terms are to ensure same current density for incident, reflected and Andreev reflected wavefunctions\cite{beenakker}. The
{Andreev reflected holes} are retro-reflected back at an angle of $\theta_A$ (in the $xy$ plane) obeying the relation $q_\rho^e Sin\theta_e = q_\rho^h Sin\theta_A$\cite{linder}.

On the other hand, for {\bf$\mu<E$} we get
\begin{align}
&\psi_N(x<0) = \frac{ e^{iq_x^{e+}x}}{\sqrt{Re[\chi_{11}]}}\left(\begin{array}{c}
  1\\
  \chi_{11}\\
  0\\
  0\\
  \end{array}\right) +\nonumber\\
  & 
   r\frac{ e^{-iq_x^{e+}x}}{\sqrt{Re[\chi_{12}]}}\left(\begin{array}{c}
  1\\
  -\chi_{12}\\
  0\\
  0\\
  \end{array}\right)
  + r_A \frac{ e^{iq_x^{h+}x}}{\sqrt{Re[\chi_{21}]}}\left(\begin{array}{c}
  0\\
  0\\
  1\\
  \chi_{21}\\
  \end{array}\right)
  \label{wvfn2}
\end{align}
with
$\chi_{21}=\chi_{22} e^{-2i\theta_A}.$

Now solving the BdG equation in the superconductor side, we get the eigenvalues for the electron-like quasiparticles (ELQ) and hole-like quasiparticles (HLQ) to be
 \begin{align}
 E_e^\pm & =\pm \sqrt{\Delta^2 + (\mu-V_s - q_z\sqrt{1+q_\rho^2})^2}\nonumber\\
 E_h^\pm & =\pm \sqrt{\Delta^2 + (\mu-V_s + q_z\sqrt{1+q_\rho^2})^2}.
 \label{scdisp}
 \end{align}
Notice that in the subgap case with $E<\Delta$, $q_\rho$ becomes imaginary that makes the mode decaying ($i.e.$, non-travelling) as it should be.
 
 The electron and hole like eigenstates come out to be
\begin{align}
\psi_e^+=\left(\begin{array}{c}
  u\\
  u\eta_1\\
  v\\
  v\eta_1\\
  \end{array}\right),~~\psi_h^-=\left(\begin{array}{c}
  v\\
  -v\eta_2\\
  u\\
  -u\eta_2\\
\end{array}\right)
\end{align}
which leads to the general wavefunction to be
\begin{align}
\psi_s(x>0)&= a\psi_s^{e+} +b \psi_s^{h-}\nonumber\\
&= a\left(\begin{array}{c}
  u\\
  u\eta_1\\
  v\\
  v\eta_1\\
  \end{array}\right)e^{ip_x^+x} +b \left(\begin{array}{c}
  v\\
  -v\eta_2\\
  u\\
  -u\eta_2\\
  \end{array}\right)e^{-ip_x^-x}
\label{wvfn3}
\end{align}
where $p_x^{+(-)}=\sqrt{[p^{+(-)}]^2-q_z^2}\cos\theta_s^{e(h)}$ with\begin{align}
  p^{\pm}&=\frac{\sqrt{((\mu-V_s)\pm \sqrt{E^2-\Delta^2})^2-q_z^2+q_z^4}}{q_z},
  \nonumber\\
  u(v)&=\sqrt{\frac{1}{2}(1+(-)\frac{\sqrt{E^2-\Delta^2}}{E})}~~~{\rm and}\nonumber\\
\eta_{1(2)}&=\sqrt{\frac{(\mu-V_s)+(-)\sqrt{E^2-\Delta^2}-q_z}{(\mu-V_s)+(-)\sqrt{E^2-\Delta^2}+q_z}}e^{i\theta_s^e(-i\theta_s^h)}.
\end{align}


\begin{figure}
  \begin{center}
   \begin{picture}(100,100)
     \put(-80,10){
        \includegraphics[width=.52\linewidth]{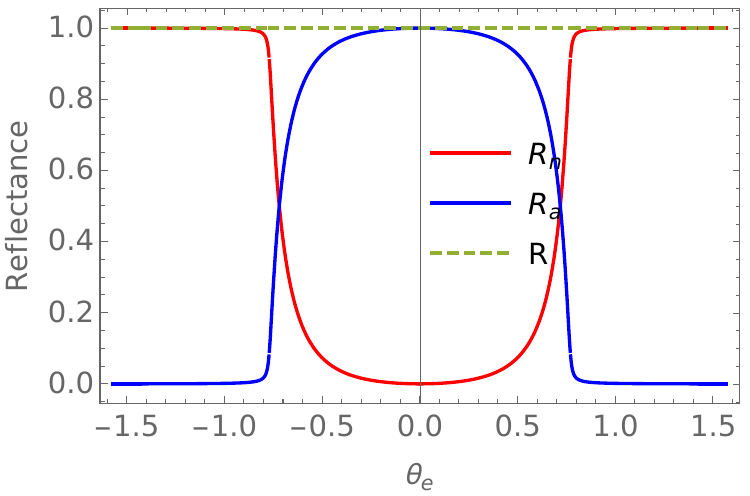}
        \includegraphics[width=.52\linewidth]{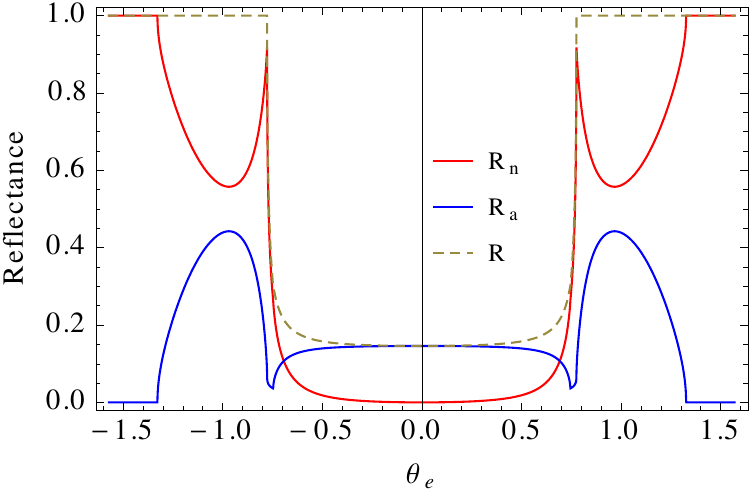}}

     \put(-80,-80){     \includegraphics[width=.52\linewidth]{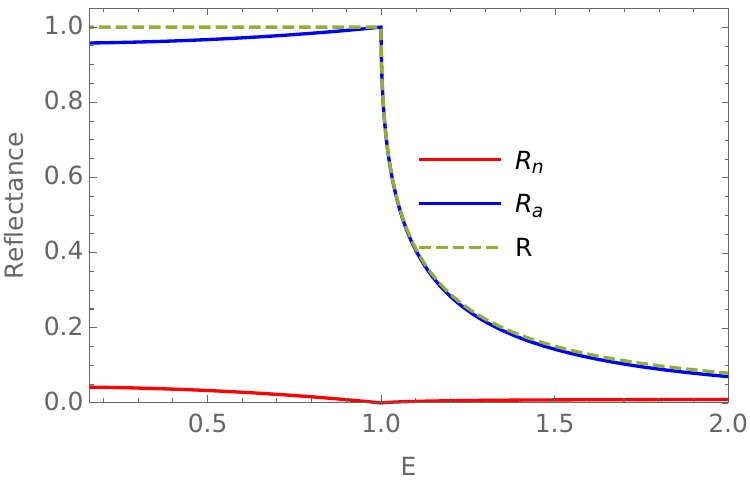}
         \includegraphics[width=.52\linewidth]{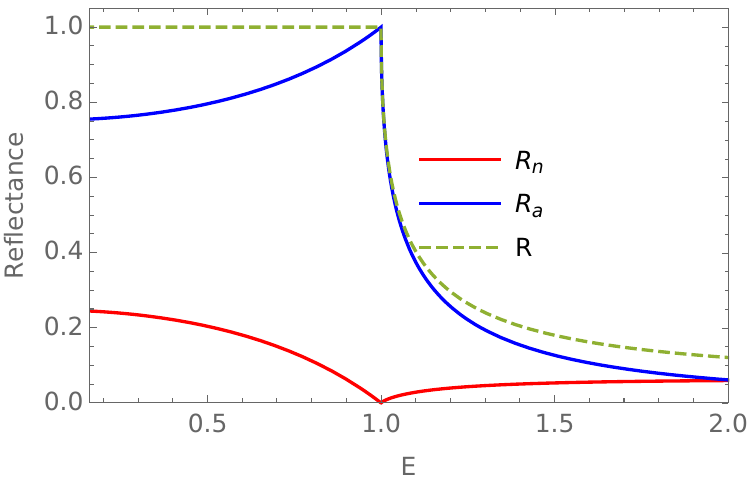}}
     \put(-80,-170){    \includegraphics[width=.52\linewidth]{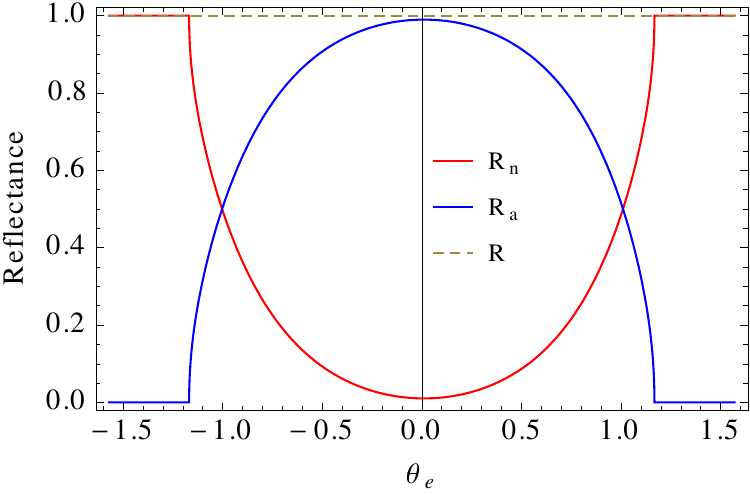}
       \includegraphics[width=.52\linewidth]{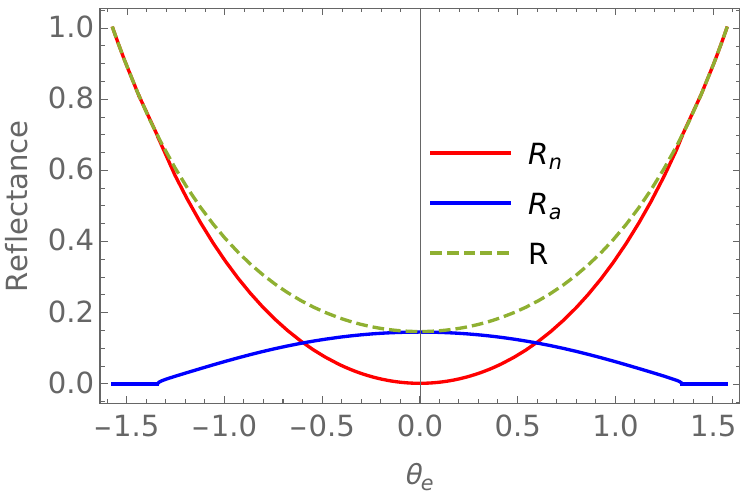}}
     
     \put(-80,-260){    \includegraphics[width=.52\linewidth]{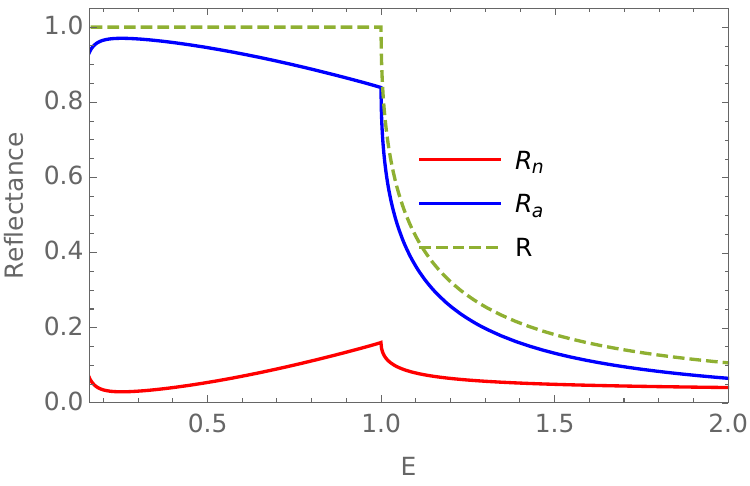}
         \includegraphics[width=.52\linewidth]{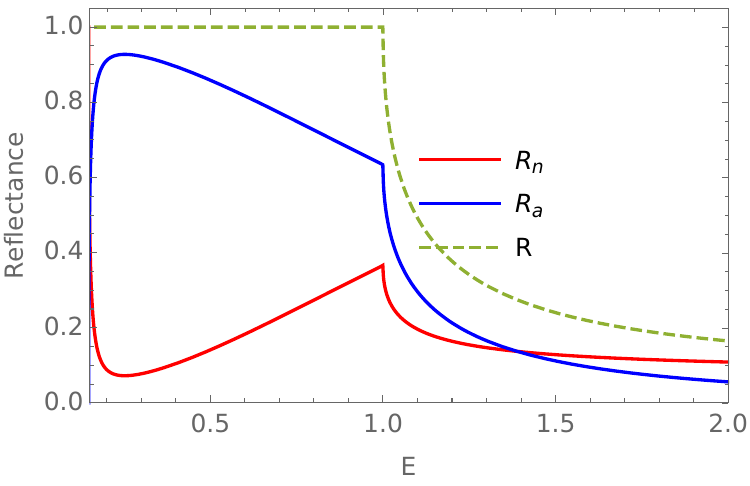}}
         \put(-45,45){(a)}
         \put(75,45){(b)}
         \put(-45,-45){(c)}
         \put(75,-45){(d)}
         \put(-55,-135){(e)}
         \put(75,-135){(f)}
         \put(-45,-225){(g)}
    \put(75,-225){(h)}
   \end{picture}  
  \end{center}
  \vskip 3.5 in
\caption{Normal and Andreev Reflectance for $E=$ (a,e) $0.5\Delta$ and (b,f) $1.5\Delta$ and $\theta_e=$ (c,g) $\pi/8$ and (d,h) $\pi/5$ with $V_s=30\Delta$ and $\mu=$ (a-d) $100\Delta$ and (e-h) $0.02\Delta$.} 
\label{refl}
    \end{figure}

Finally matching the wave functions at the boundary with $\psi_N(x=0^-)=\psi(x=0^+)$, we solve for the four unknowns $r,~r_A,~a$ and $b$.

Typical plots for normal reflectance $R_n=r^2$ and Andreev reflectance $R_a=r_A^2$ are shown in Fig.\ref{refl} for different direction and energy (both subgap and supergap) of the incident electrons. Notice that for normal incidences, $R_a=1$ only in the subgap case with $E<\Delta$ whereas in the supergap region, $R_a$ decreases with an increase in $E$.

For $\mu>>\Delta$, $R_a$ decreases steadily to zero as $\theta_e$ is increased gradually from $0$ in the subgap cases. However for $E>\Delta$, such steady decrease is halted as $R_a$ shows a sudden upturn at $\theta_e\sim\pi/4$ which continues till $R_a$ maximizes at $\theta_e\sim\pi/3$ and then decreases to zero at $\theta_e\sim\theta_c=\sin^{-1}[\frac{q_N^h}{q_N^e}]$. {For $\theta_e\gtrsim\pi/4$, $\theta_s^e$ and $\theta_s^h$ becomes complex turning quasiparticle modes in the SC side decaying (and hence only Andreev transport remains present in the quantum tunnelling). Interestingly, this switching off of the quasiparticle transport makes the scenario similar to subgap cases and the Andreev transport become stronger signalling the sudden increase in $R_a$.}

 For $\mu<<\Delta$, the decrease in $R_a$ with $\theta_e$ maintains its monotony. Interestingly, both $R_n$ and $R_a$ behavior reverses around $E\sim q_z\pm\mu$. For higher energies, an increase in $E$ causes $R_a$ to decrease monotonically till $E=\Delta$. Beyond that it registers a sharper decay due to resistive losses.
 
 The reflection probabilities behave as symmetric functions of the angle $\theta_e$ as long as $\chi_{11}^0$ and $\chi_{22}^0$ are real (see Eq.\ref{eq4}). Fig.\ref{refl}(a) shows that Andreev reflection $R_a=1$ and normal reflection $R_n=0$ when $E$=0.5 and $\mu=100\Delta$ for $\theta_e=0$. As we increase the incidence angle $\theta_e$ from 0 to $\pi/2$, Andreev reflection decreases gradually and normal reflection increases. At $\theta_e=\pi/4$, $R_n$ and $R_a$ both are equal to 0.5. With $E>\Delta$ (as given in Fig.\ref{refl}(b) for $E=1.5$), $R_a$ gets reduced from its maximum limit of unity even for normal incidence along $xy$ plane, which further decreases with increase in $\theta_e$.
 For a small $\mu=.02\Delta$ as well (see Fig.\ref{refl}(e)), we see the Andreev reflection to be maximum at $R_a=1$ and normal reflection $R_n=0$ when $E=0.5$. As increase the incidence angle $\theta_e$ from 0 to $\pi/2$, AR decreases gradually and normal reflection increases. At a particular angle $\theta_e\sim\pi/3$, Andreev reflection and normal reflection both are equal to 0.5. In the supergap regime ($E=1.5$ in Fig.\ref{refl}(f), quasiparticle transport in the SC side results in $R_a+R_n<1$ for almost the whole range of $\theta_e:~(0,\pi/2)$.

\begin{figure}
  \includegraphics[width=1.1\linewidth]{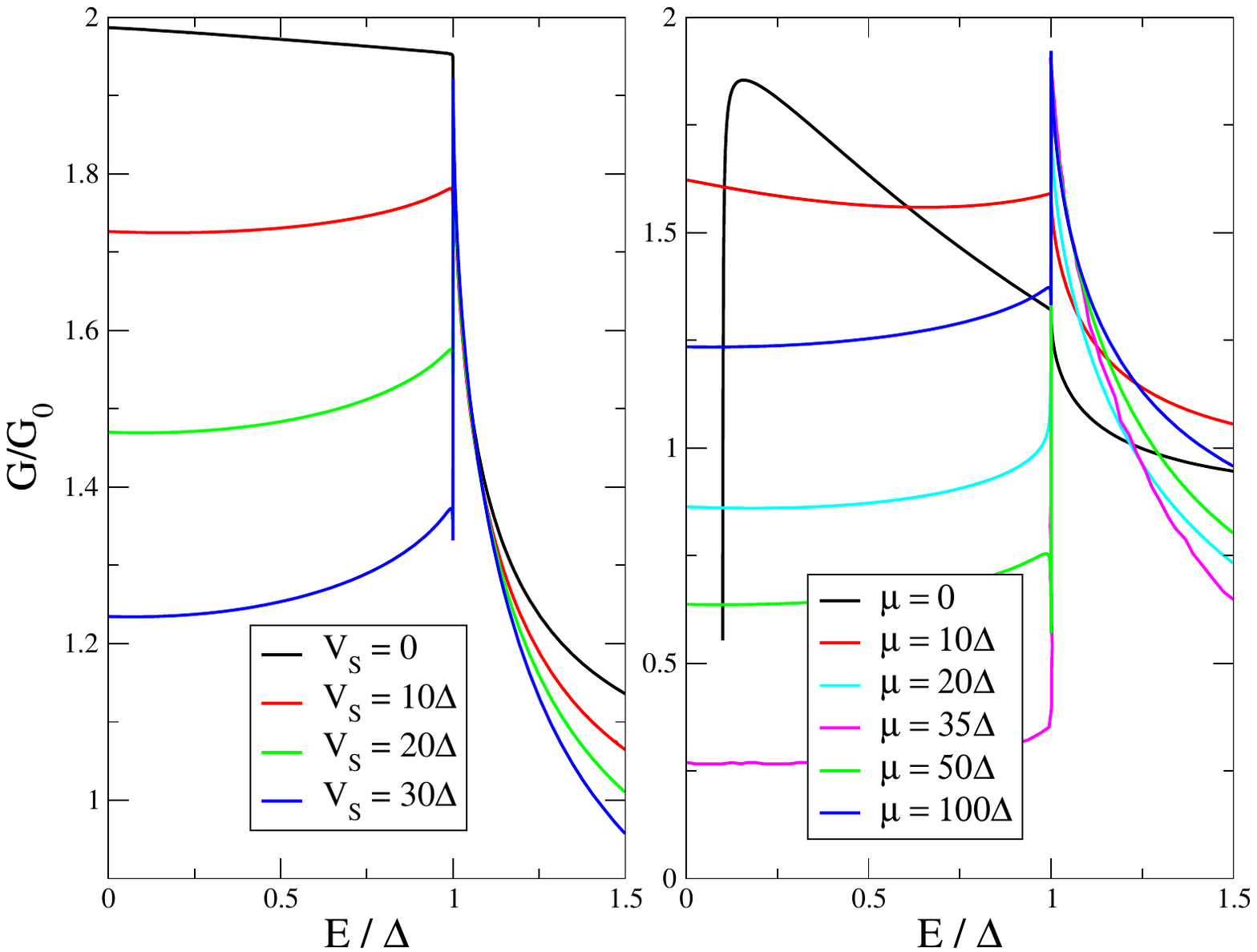}
  \put(-170,150){(a)}
  \put(-50,150){(b)}
       \caption{Tunnelling Conductance $G/G_0$ for $q_z=0.1$. We consider (a) $\mu=100\Delta$ and (b) $V_s/\Delta=$ 30 for different values of $V_S$ and $\mu$ respectively.} 
\label{cond}
\end{figure}

We can also evaluate the differential tunnelling conductance of the NSSM-SC junction, for fixed $q_z$, using Blonder-Tinkham-Klapwijk formula\cite{btk,beenakker} as
\begin{equation}G/G_0=\int_0^{\pi/2}(1+R_a-R_n)\cos\theta_e d\theta_e.
\end{equation}
where $G_0$ denotes the ballistic conductance of the NSSM \cite{debu}.
In Fig.\ref{cond}, we show that $E$ dependence of the conductance. Let's consider the cases for large $\mu$ first and examine the results in Fig.\ref{cond}(a).
$V_s=0$ implies absence of normal reflections at $E=0$. From there an increase in $E$ in the subgap case can only cause $R_a$ to decrease as the chances of creating Cooper pairs in the SC side reduces for smaller ($\Delta-E$). This is manifested by a slow decrease in $G/G_0$ till $E=\Delta$. At $E=0$, an increase in $V_s$ from zero, however, causes $R_n$ to increase thereby reducing $G/G_0$. As $E$ is increased from there, $R_a~(R_n)$ starts increasing (decreasing) causing $G/G_0$ to increase till $E=\Delta$. For $E>\Delta$, quasiparticle states become available in the SC side and $G$ shows a resistive decay with $E$ in those limits.

Next we decrease $\mu$ as seen in Fig.\ref{cond}(b). The difference $\mu-V_S$ plays a significant role in the conductance as one can understand from the dispersion relation Eq.\ref{scdisp}. A decrease in $\mu$ (and hence in $\mu-V_S$) causes effective gap in the SC spectrum (at the fixed $q_z=0.1$) to decrease which in turn reduces $R_a$ and consequently the subgap conductivity till $\mu=V_S$. 
{For $\mu\sim V_s$, Andreev reflection occurs mostly for normal incidences and $G$ becomes very small at low energies, though it again increases} upto the maximum at $E=\Delta$. With further decrease in $\mu$, the subgap conductance starts increasing again. However, no conductivity can be expected for $E<q_z$ as it turns $q_\rho$ imaginary.


\section{Tunnelling in presence of irradiation}
{An irradiation introduces electromagnetic fields to the problem and the canonical momentum of the charge carriers changes as per the Peierl's substitution: $p\rightarrow p-e\mathcal{A},~\mathcal{A}$ being the corresponding vector potential. Given the time periodic nature of $\mathcal{A}$, an effective Hamiltonian can be designed at least within some approximation. For large frequency of the field\cite{cmnt}, an high-frequency expansion is employed and an effective Floquet Hamiltonian is obtained keeping first few terms ($i.e.,$ the prominent ones) of the expansion\cite{eckardt}.
  }

Ref.\cite{jpcm} has shown how different Floquet Hamiltonians ($H_F$'s) can be obtained by irradiating a NSSM (like one considered here) for different polarization of radiation. This being in junction with superconductors can offer many interesting transport features that we are going to explore now.

\subsection{Floquet system for circular polarization}
If we consider a circularly polarized light of angular frequency $\omega$, we get a $H_F$ given by
\begin{align}\label{cpe}
H_F&=q_z(q_x\sigma_x+q_y\sigma_y)+q_z\sigma_z +\frac{(eE_0q_z)^2}{\hbar^3\omega^3}\sigma_z\nonumber\\
&=q_xq_z\sigma_x+q_yq_z\sigma_y+q_z(1+q_1q_z)\sigma_z
\end{align}
with $q_1=q_0^3=\frac{1}{\hbar^3\omega^3}$ and $(eE_0)^2=1$. {Notice that this Hamiltonian indicates presence of a Weyl point (more precisely a multi-Weyl point\cite{jpcm}) at $(0,0,-q_1^{-1})$ in addition to the NS at $q_z=0$ and that makes the spectral and hence transport features of irradiated model distinctive compared to the unirradiated model.}\\
Putting Eq.\ref{cpe} in the BdG equations for transport along $x$ leads to the quasiparticle dispersions for electrons and holes as
\begin{align}
E_e^\pm &= \pm q_z\sqrt{(1+q_1q_z)^2+q_\rho^2}-\mu
\nonumber\\
E_h^\pm &= \pm q_z\sqrt{(1+q_1q_z)^2+q_\rho^2}+\mu
\end{align}\\
Assuming again an electron with the wave vector $q_x^e$ injected from $-\hat{x}$ direction in the 
  NSSM, the wave function for $\mu>E$ and $\mu<E$ can be obtained as Eq.\ref{wvfn1} and \ref{wvfn2} respectively with modification given by
  \begin{figure}
    \begin{center}
   \begin{picture}(100,100)
     \put(-80,10){
        \includegraphics[width=.52\linewidth]{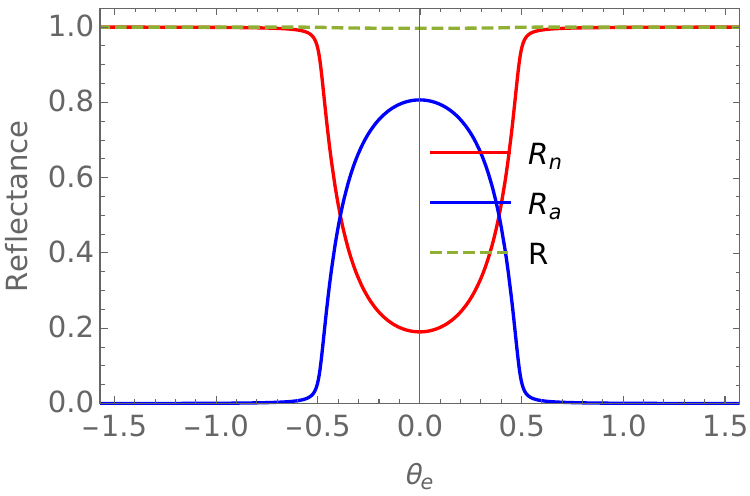}
        \includegraphics[width=.52\linewidth]{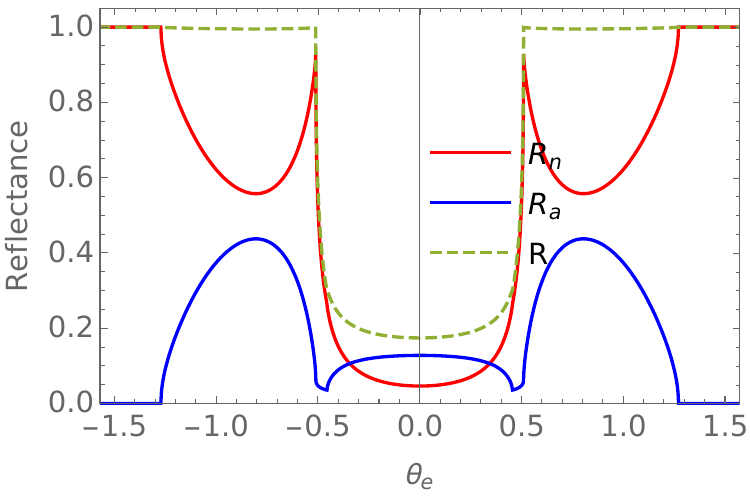}}
     
      \put(-80,-80){   \includegraphics[width=.52\linewidth]{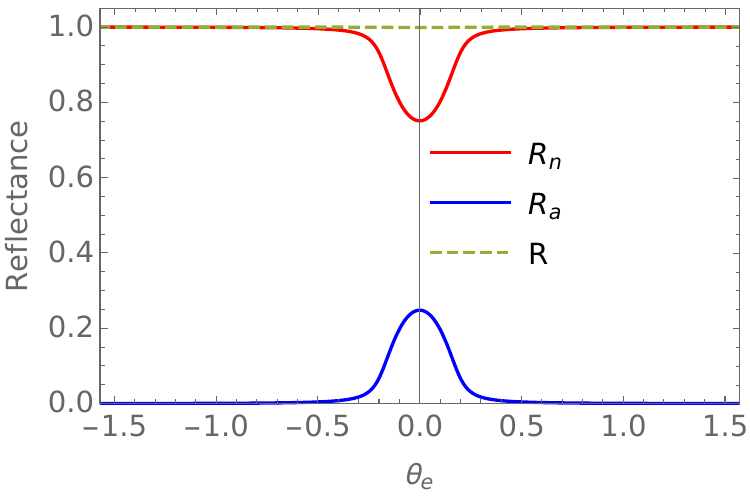}
         \includegraphics[width=.52\linewidth]{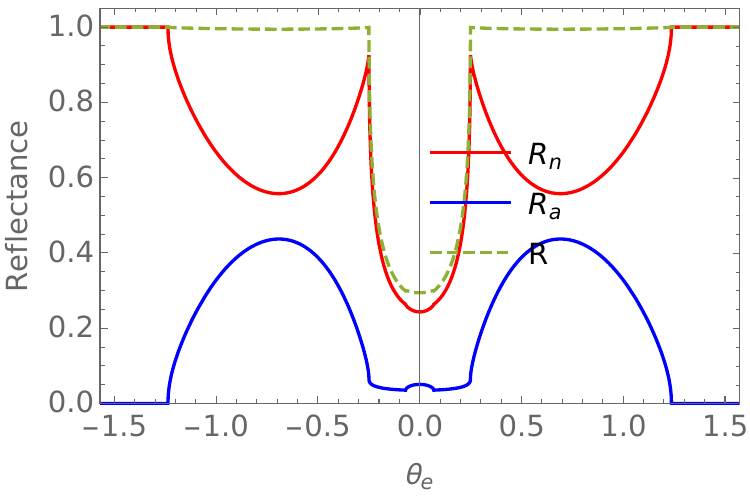}}
      \put(-80,-170){   \includegraphics[width=.52\linewidth]{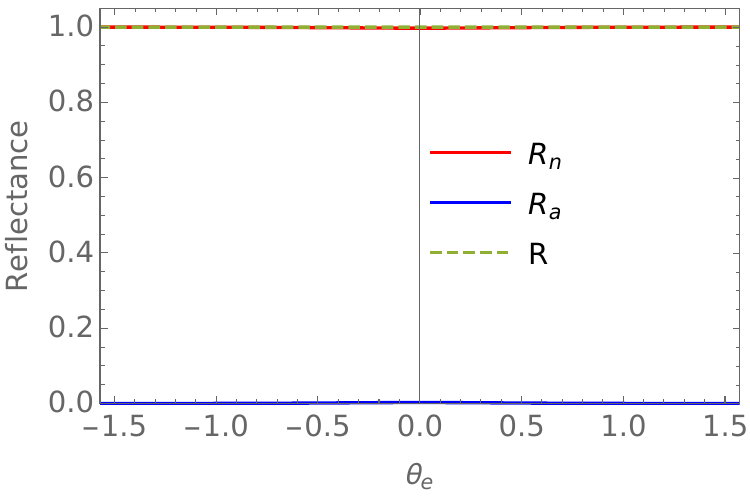}
        \includegraphics[width=.52\linewidth]{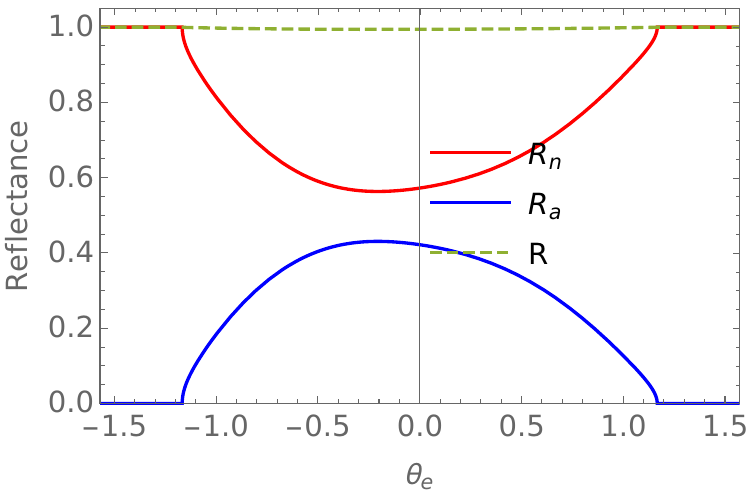}}
     \put(-80,-260){    \includegraphics[width=.52\linewidth]{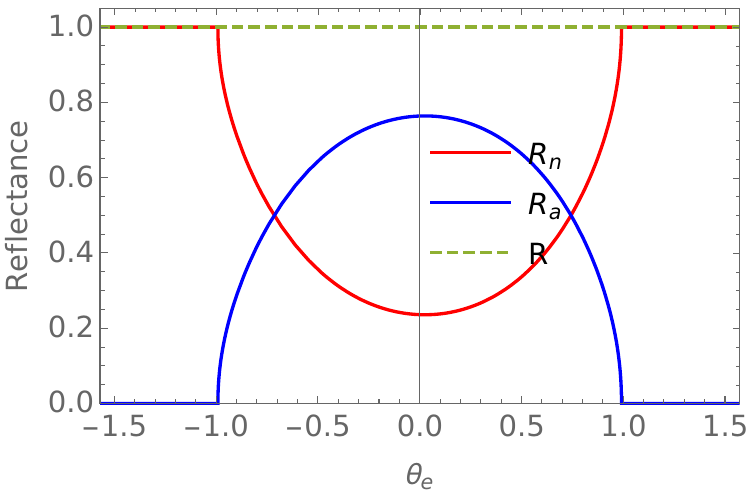}
         \includegraphics[width=.52\linewidth]{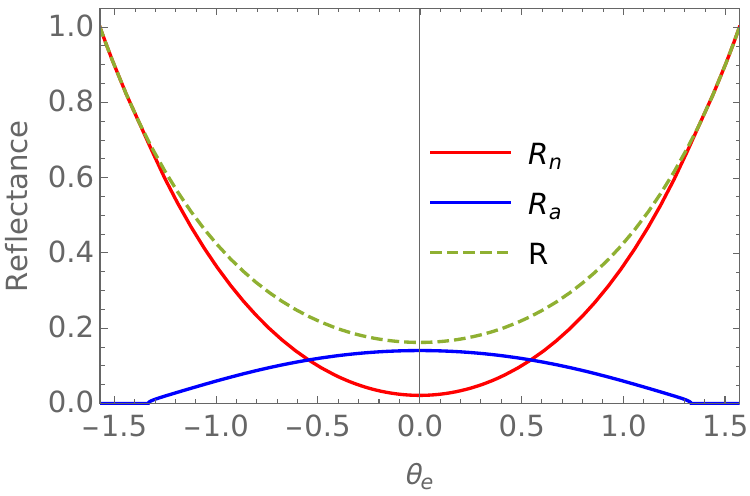}}
         \put(-45,55){(a)}
         \put(75,55){(b)}
         \put(-45,-45){(c)}
         \put(75,-45){(d)}
         \put(-45,-115){(e)}
         \put(75,-115){(f)}
         \put(-45,-220){(g)}
    \put(75,-220){(h)}
   \end{picture}  
  \end{center}
  \vskip 3.5 in
\caption{Normal and Andreev Reflectance for $E=$ (a,c,e,g) $0.5\Delta$ and (b,d,f,h) $1.5\Delta$ with $V_s=30\Delta$ and $\mu=$ (a-f) $100\Delta$ and (g-h) $0.02\Delta$. We consider $q_0=18$ (a,b), $19$ (c,d), $20$ (e,f) and $3$ (g,h) respectively.} 
\label{reflcp}
    \end{figure}

\begin{align}
\chi_{11}=\sqrt{\frac{(E +\mu)-q_z(1+q_1q_z)}{(E +\mu)+q_z(1+q_1q_z)}} e^{i\theta_e},~~\chi_{12}=\chi_{11} e^{-2i\theta_e}\nonumber\\
\chi_{22}=\sqrt{\frac{(E -\mu)+q_z(1+q_1q_z)}{(E-\mu)-q_z(1+q_1q_z)}} e^{i\theta_A}
,~~\chi_{21}=\chi_{22} e^{-2i\theta_A}.
\end{align}
Here 
$\theta_A$'s are again calculated from the relation $q_\rho^e sin\theta_e = q_\rho^h sin\theta_A$\cite{linder}.

Then we look at the SC side where the ELQ and HLQ dispersions turn out to be
 \begin{align}
 E_e^\pm &=\pm \sqrt{\Delta^2 + ((\mu-V_s) - q_z\sqrt{(1+q_1q_z)^2+q_\rho^2})^2}\nonumber\\
 E_h^\pm &=\pm \sqrt{\Delta^2 + ((\mu-V_s) + q_z\sqrt{(1+q_1q_z)^2+q_\rho^2})^2}
 \end{align}
 \begin{figure}
   \includegraphics[width=\linewidth]{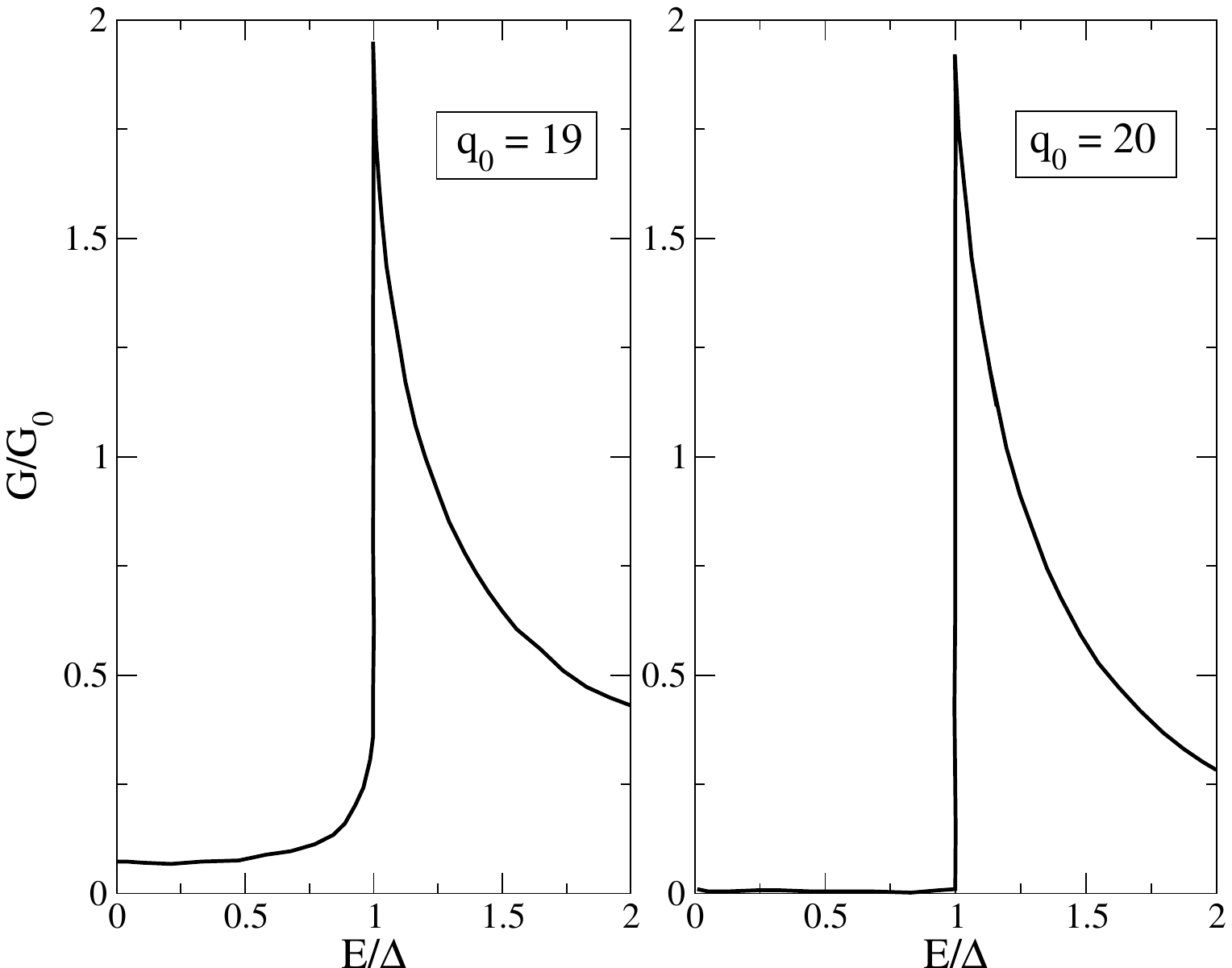}
   \put(-200,100){(a)}
  \put(-100,100){(b)}
       \caption{Tunnelling Conductance $G/G_0$ for $q_z=0.1$. We consider $q_0=$ (a) 19 and (b) 20 respectively for $\mu=100\Delta$ and $V_S=30\Delta$.} 
\label{cond2}
\end{figure}
 The Wave functions are again given by Eq.\ref{wvfn3} with
\begin{align}
\eta_1=\sqrt{\frac{(\mu-V_s)+\sqrt{E^2-\Delta^2}-q_z(1+q_1q_z)}{(\mu-V_s)+\sqrt{E^2-\Delta^2}+q_z(1+q_1q_z)}}e^{i\theta_s^e}
\end{align} 
\begin{align}
\eta_2=\sqrt{\frac{(\mu-V_s)-\sqrt{E^2-\Delta^2}-q_z(1+q_1q_z)}{(\mu-V_s)-\sqrt{E^2-\Delta^2}+q_z(1+q_1q_z)}}e^{-i\theta_s^h}
\end{align}
and wavefunction match at the boundary lead us to
the reflection coefficient(r) and Andreev reflection coefficient$(r_A)$.

Fig.\ref{reflcp} shows the angular variation of reflectivities for different $q_0$ values. {Note that for small $q_0$, the Weyl point present in the spectrum remains far from the NS (as for Weyl point $q_z=-q_1^{-1}>>0$ for small $q_0$) and low energy transport barely feels its effect. Particularly,} in the limit of $\mu>>\Delta$ all the variations are similar with and without irradiation for the small value of $q_0(=\frac{1}{\hbar\omega})$ upto $0<q_0\lesssim 17$. But a larger $q_0$ can show different behavior (as seen in Fig.\ref{reflcp}) where for small $\theta_e$, subgap Andreev reflectance gets reduced from its maximum of unity and interestingly supergap Andreev reflectance first shows a decrease (going from $q_0=18$ to $19$) followed by an increase (going from $q_0=19$ to $20$) finally indicating no quasiparticle transmission (because $R_n+R_a=1$) on the SC side. The asymmetry with respect to $\theta_e$ also become discernible for supergap cases (see Fig.\ref{reflcp}(f)). Even in the limit of $\mu<<\Delta$ subgap Andreev reflectance gets reduced (from unity for $\theta_e=0$) due to irradiation.

In Fig.\ref{cond2}, we show two typical tunneling conductance plots of our irradiated system for $\mu=100$ and $V_s=30$. Due to heavy reduction of subgap Andreev reflectance, we see that $G$ becomes very small for $E<\Delta$ in the case for $q_0=19$, which almost gets perished in the case of $q_0=20$. Also the resistive decay of $G$ in the supergap regime becomes faster for $q_0=20$ compared to that for $q_0=19$.

\subsection{Floquet system for linear polarization}

In case of a linearly polarized irradiation, the Floquet Hamiltonian can become\cite{jpcm}
\begin{align}\label{lpe}
  H&=q_xq_z\sigma_x+q_y(q_z-q_2q_z^3)\sigma_y+q_z(q_2q_z-q_z^3)\sigma_z\nonumber\\
&=q_xq_z\sigma_x+q_yq_z(1-q_2q_z^2)\sigma_y+q_z(1-q_2q_z^2)\sigma_z
\end{align}
where $q_2=q_0^4/2=\frac{1}{2(\hbar\omega)^4}$ when $(eE_0)^2=1$.
{Unlike Eq.\ref{cpe}, this Eq.\ref{lpe} distinguishes between $x$ and $y$ direction and not symmetric about $x\leftrightarrow y$ conversion. Moreover, in this case the prefactor $q_2\sim\omega^{-4}$ as opposed to the prefactor $q_1\sim\omega^{-3}$ in Eq.\ref{cpe}. Furthermore, other than the NS at $q_z=0$ this also constitutes a nodal line at $(0,q_y,q_2^{-1})$. All these differences of Hamiltonian (\ref{lpe}) with Hamiltonian equations (\ref{eq1}) or (\ref{cpe}) reflect in its transport characteristics. Again, similar to the previous case with circular polarization, here also for a small $q_0$, the nodal line stays very far apart from the NS and low energy transport hardly sees the presence of it.
  }

Corresponding to the quasiparticle dispersions of electron and holes within the NSSM look like:
\begin{align}
E_e^\pm &= \pm q_z\sqrt{q_x^2+(1+q_y^2)(1-q_2q_z^2)^2}-\mu
\nonumber\\
E_h^\pm &= \pm q_z\sqrt{q_x^2+(1+q_y^2)(1-q_2q_z^2)^2}+\mu
\end{align}\\

For simplicity we consider $q_y=0$ and keep the transport in he $x-z$ plane alone.
Assuming an electron with the wave vector $q_x^e$ to be injected towards the junction, the wave functions can be obtained again similarly with parameters modified to be
\begin{align}
\chi_{11}=\sqrt{\frac{(E +\mu)-q_z(1-q_2q_z^2)}{(E +\mu)+q_z(1-q_2q_z^2)}},~~\chi_{12}=\chi_{11},
\nonumber\\
\chi_{22}=\sqrt{\frac{(E -\mu)+q_z(1-q_2q_z^2)}{(E -\mu)-q_z(1-q_2q_z^2)}},~~\chi_{21}=\chi_{22}.
\end{align}

 The dispersions for ELQ and HLQ in the superconducting side look like
 \begin{align}
 E_e^\pm &=\pm \sqrt{\Delta^2 + ((\mu-V_s) - q_z\sqrt{q_x^2+(1+q_y^2)(1-q_2q_z^2)^2})^2}\nonumber\\
 E_h^\pm &=\pm \sqrt{\Delta^2 + ((\mu-V_s) + q_z\sqrt{q_x^2+(1+q_y^2)(1-q_2q_z^2)^2})^2}
 \end{align}
 We again consider $q_y=0$ for simplicity.
 
Wavefunctions are obtained with parameters modified as
\begin{align}
\eta_1=\sqrt{\frac{(\mu-V_s)+\sqrt{E^2-\Delta^2}-q_z(1-q_2q_z^2)}{(\mu-V_s)+\sqrt{E^2-\Delta^2}+q_z(1-q_2q_z^2)}}
\end{align} 
\begin{align}
\eta_2=\sqrt{\frac{(\mu-V_s)-\sqrt{E^2-\Delta^2}-q_z(1-q_2q_z^2)}{(\mu-V_s)-\sqrt{E^2-\Delta^2}+q_z(1-q_2q_z^2)}}
\end{align}\\

Wave function match at the boundary gives the reflection coefficient(r) and Andreev reflection coefficient$(r_A)$'s.

\begin{figure}
  \begin{center}
    \vskip 1.43 in
    \begin{picture}(100,100)
      \put(-90,0){
    \includegraphics[width=1.2\linewidth]{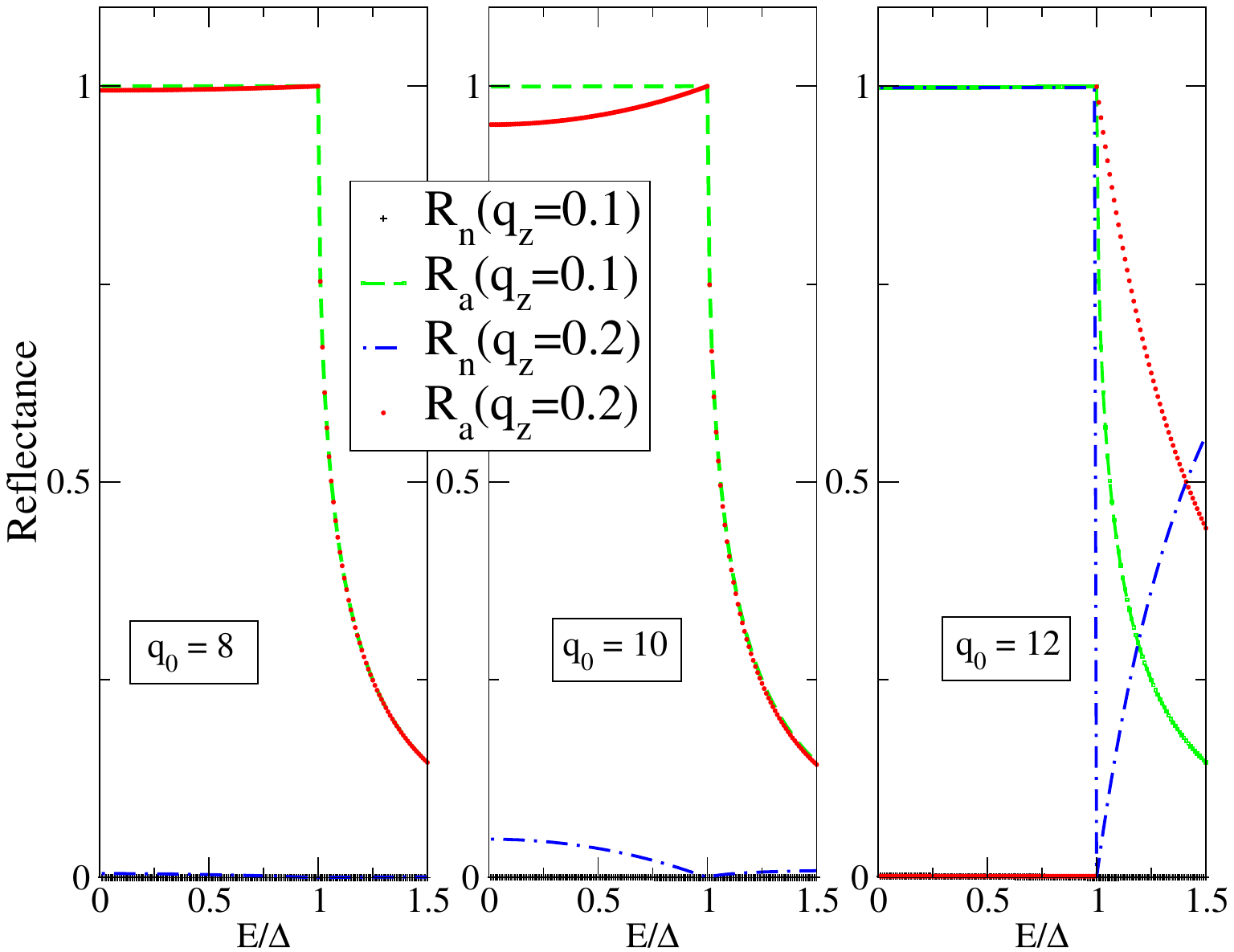}}
   \put(-40,80){(a)}
   \put(50,80){(b)}
    \put(120,80){(c)}
 \end{picture}
  \end{center}
  \vskip -.1 in
    \caption{Normal and Andreev Reflectance for $q_{0}=$ (a) 8, (b) 10 and (c) 12 respectively for $V_s=30\Delta$ and $\mu=100\Delta$ and $q_z=0.1$ and $0.2$.}
\label{linear}
\end{figure}

In this case, for $\mu>E$ all the plots variation are same with and without irradiation for the small value of $q_0(=\frac{1}{\hbar\omega})$, for $0<q_0<2$.

 In Fig.\ref{linear}, reflection probabilities are plotted as functions of the incident energy with different linear polarization parameter $q_0=8$, $q_0=10$ and $q_0=12$ for two different $q_z$ values with constant $\mu=100$, $V_s=30$. For $q_z=0.1$, normal reflectance remains zero for small incident energies ($E$). Andreev reflectance stays at unity for subgap cases and decay gradually for $E>\Delta$. But for a higher value of $q_z=0.2$, $R_n$ and $R_a$ are not constant for $0<E<\Delta$. Starting from unity, the value of $R_a$ decreases with $q_0$ at $E=0$ which then increases to become $R_a=1$ at $E=\Delta$. Beyond that one can notice monotonic decay of $R_a$ with increasing $E$.  Normal reflectance $R_n$ start growing in the subgap case as $q_0$ is increased which shows the maximum vale of unity for $q_0=12$. There in the supergap regime also $R_n$ keeps increasing with the value of $E$.

 \section{Discussion and Summary}
 Our work presents charge transport characteristics for quantum tunneling through a SC-NSSM junction, both in the presence and absence of light irradiation. In our model, NS appears for $q_z=0$ and we probe the reflectance and conductances considering small fixed values of $q_z$ close to the nodal surface. For junction interface perpendicular to NS that we consider here,
 we find single SAR or RAR (unlike double reflections obtained for nodal line semimetals\cite{prb101}) to occur during charge transport. RAR dominates for $\mu>>\Delta$ and SAR dominates for $\mu<<\Delta$ which is similar to behavior observed in Graphene-SC junctions \cite{beenakker}. {However, due to the typical dispersion characteristics of the NSSM, many distinguishing variation of reflectance and tunneling conductances are witnessed in this case. For example, we witness an unconventional rise and fall of supergap Andreev reflectance at intermediate incident angles for large $\mu$ values ($e.g.,$ see the behavior within $\theta_c>|\theta_e|\gtrsim\pi/4$ as shown in Fig.\ref{refl}b). This also marks vanishing of the quasiparticle transport across the junction. Furthermore for small $\mu$, a change in the behavior of reflectance is observed when $E$ surpasses the value of $q_z$ (see Fig.\ref{refl}g-h). Also for $\mu\sim V_S$, tunneling conductance declines at subgap energies as AR occurs only for normal incidence. All such nontrivial variations make the study of transport through a NSSM-SC junction very special and provide ample scopes for tuning subgap or supergap conductivities as per requirements with the variation of the parameters like $\theta_e$, $E$, $V_S$, $\Delta$ or $\mu$.}
 
 Then we also consider the dynamics of these transport features by the application of irradiation via linear and circular polarizations. These indicate even richer variation of transport properties. On increasing the irradiation parameter $q_0(=\frac{1}{\hbar\omega})$, reflection probabilities change drastically. {In the Floquet Hamiltonian for the case of circular polarization, a large $q_0$ let the NS and Weyl point of the spectrum stay close to each other and our calculations indicate nearly depleted subgap Andreev conductivities in that limit (see Fig.5). Similarly for the case of linear polarization, a large $q_0$ let the NS stay close to the nodal line of the Floquet system. There we find normal subgap reflectance to start from zero at very small $q_z$ values and then increases with $q_z$ as well as with $q_0$ {(see Fig.6)}. Interestingly such coexistence of nodal surface and nodal line has also been observed from angle resolved photo-emission spectroscopy (ARPES) results in $ZrSiS$\cite{zrsis}.} We find that at normal incidence $\theta_e=0$, $R_a=1$ for $E<\Delta$ in the absence of light irradiation while with irradiation, $R_a$ decreases with increase in $q_0$, both for the case of circular and linear polarization {(see Fig.4a,c,e and Fig.6)} with $R_n$ increasing even for $E>\Delta$ {(see Fig.4b,d,f and Fig.6)}. Interestingly for large $q_0$, sum of $R_n$ and $R_a$ remains unity not only in the subgap region but also in the supergap region (for $E$ not very large) indicating no quasiparticle transport within the SC side. However, Andreev reflection still causes the tunneling conductance to decay in the supergap region.

 The reported exotic variations of $R_n$, $R_a$ and $G$ for NSSM based SN junction {can provide many-fold tunabilities} in designing electronic devices like superconducting LED\cite{led}, solar cells\cite{solar} or supercurrent diodes\cite{diode}. {Also} the tuning of these transport behaviors are much easier to implement if done using irradiation than by straining\cite{ali} the system or opening the spectral gap\cite{majidi} artificially. 
 One can {even investigate} the entanglement of the transport carriers in the SC junction or their Floquet version coming due to irradiation\cite{banasri} for its possible use in the field of quantum information processing\cite{jozsa}.  Such variations in reflectance and conductance can well be examined in a properly replicated cold atom set-up on optical lattices\cite{gross} {and one can even perform additional tuning of the parameters for further experimentation.}
 
{ In this regard we may add that recently, ab-initio calculations found topological nodal surfaces in binary compounds $Sr_5X_3~(X=As,Sb,Bi)$\cite{dft2} or Polyhydride material $ScH_3$ and $LuH_3$\cite{dft}. These were already experimentally synthesized with the later ones even showing superconductivity at both ambient and high pressures\cite{expt}. Thus one no more necessarily requires proximity to an external superconductor to induce superconductivity in nodal surfaces like that considered in our work. These materials can become candidates for topological superconductivity\cite{dft} that can host exotic zero energy Majorana bound states\cite{sdsharma}.
 Interestingly, three dimensional higher order NSSM with stacks of 2D anisotropic Su-Schrieffer-Heeger lattice exhibiting hinge arcs connecting projected nodal surfaces has also been proposed (and demonstrated in acoustics as well)\cite{qiu} in recent time.}

\section*{Acknowledgements}
SK thanks D. Sinha for fruitful discussions and acknowledges financial support from DST-SERB (now called ANRF), Government of India via grant no. CRG/2022/002781.


\begin{thebibliography}{99}

  
\bibitem{wen} X.-G. Wen, Rev. Mod. Phys.{\bf 89}, 041004 (2017).
\bibitem{armitage} N. P. Armitage $et.~al.$, Rev. Mod. Phys.{\bf 90}, 015001 (2018).
\bibitem{skrev} {S. Kar, A. Jayannavar, Asian Jour. of Res. and Rev. in Phys., {\bf 4(1)}, 34-45 (2021).}
  \bibitem{fang} C. Fang, Chin. Phys. B{\bf 25}, 117106 (2016).

\bibitem{wu} W. Wu $et~al.$, Phys. Rev. B{\bf 97}, 115125 (2018).
\bibitem{zhong} C. Zhong $et~al.$, Nanoscale {\bf 8}, 7232 (2016).
     \bibitem{turker} O. Turker $et~al.$, Phys. Rev. B{\bf 97}, 075120 (2018).
\bibitem{xiao} M. Xiao $et~al.$, arXiv:1709.02363 (2017); M. Xiao $et~al.$, Sci. Adv.{\bf 6}, eaav2360 (2020).
\bibitem{yang} Y. Yang $et~al.$, Nat. com. {\bf 10}, 5185 (2019).
   \bibitem{jpcm} B. Pandit $et~al.$, Jour. Phys. Cond.-Mat.{\bf 37}, 075601 (2025).
     \bibitem{liang} Q.-F. Liang $et~al.$, Phys. Rev. B {\bf 93}, 085427 (2016).
  \bibitem{furusaki} A.Furusaki, Science Bulletin {\bf 62}, 788-794 (2017).
   \bibitem{zhao} Y. X. Zhao $et~al.$, Phys. Rev. B{\bf 94}, 195109 (2016).
 \bibitem{andreev}    A. F. Andreev, Sov. Phys. JETP {\bf 19}, 1228 (1964).
 \bibitem{beenakker} C. W. J. Beenakker, Phys. Rev. Lett.{\bf 97}, 067007 (2006).
  {\bibitem{btk} G.E. Blonder, M. Tinkham, and T. M. Klapwijk, Phys. Rev. B{\bf 25}, 4515 (1982).} 
\bibitem{silicene} J. Linder $et~al.$, Phys. Rev. B{\bf 89}, 020504(R) (2014).
   \bibitem{mos2} L. Majidi $et~al.$, Phys. Rev. B{\bf 89}, 045413 (2014).
   \bibitem{phosphorene} J. Linder $et~al.$, Phys. Rev. B{\bf 95}, 144515 (2017).
  \bibitem{wsm} S. Uchida $et~al.$, J. Phys. Soc. Jpn.{\bf 83}, 064711 (2014).
     \bibitem{prb101} Q. Cheng $et~al.$, Phys. Rev. B{\bf 101}, 094508 (2020).
     \bibitem{eckardt} A. Eckardt $et~al.$, New Jour. Phys. {\bf 17}, 093039 (2015).
     {\bibitem{cmnt} In the small frequency limit instead, one can use techniques like Adiabatic-Impulse approximations to study the dynamics of the problem. This approach is outlined in many articles like Rev.Mod.Phys.(83)863, Phys.Rev.B(94)075130 or Phys.Rev.B(95)085141.
       However, in this work we only kept ourselves confined within the high frequency limit.}
    \bibitem{moessner} J. Cayssol $et~al.$, Phys. Stat. Solidi RRL{\bf 7}, No.1-2, 101 (2013).
\bibitem{debu} D. Sinha $et~al.$, Cur. App. Phys.{\bf 18}, 1087 (2018).
 \bibitem{cheng} S.-G. Cheng $et~al.$, Phys. Rev. Lett.{\bf 103}, 167003 (2009).
 \bibitem{xing} Y. Xing $et~al.$, Phys. Rev. B 83, 205418 (2011).
 \bibitem{linder} J. Linder $et~al.$, Phys. Rev. B{\bf 77}, 064507 (2008).
   { \bibitem{zrsis} B. -B. Fu $et~al.$, Sci. Adv. {\bf 5}, eaau6459 (2019).}
       \bibitem{led} S. S. Mou $et~al.$, JSTQE, 2346617 (2014).
    \bibitem{solar} L. M. S.-Gutierrez $et~al.$, J. Phys. Chem. C{\bf 128}(45), 19329 (2024).    
 \bibitem{diode} J. J. He $et~al.$, Nat. Com.{\bf 14}, 3330 (2023).

\bibitem{ali} M. Alidoust, J. Linder, Phys. Rev. B 84, 035407 (2011).
\bibitem{majidi} L. Majidi, M. Zareyan, Phys. Rev. B 86, 075443 (2012).
  \bibitem{banasri} S. Kar $et~al.$, Phys. Rev. B{\bf 98}, 245119 (2018).
  \bibitem{jozsa} R. Jozsa, and N. Liden, Roy. Soc. Proc. A, Vol.{\bf 459}, Issue 2036,  1097 (2003).
  \bibitem{gross} C. Gross, and I. Bloch, Science, Vol.{\bf 357}, Issue 6355, 995-1001 (2017).
    {\bibitem{dft2} M. R. Khan $et~al.$, Phys. Rev. B 105, 245152 (2022).
      \bibitem{dft} A. Sufiyan, J. A. Larsson, ACS Omega 8, 9607 (2023).
\bibitem{expt} M. Shao $et~al.$, Inorg. Chem. 60, 15330 (2021).
 \bibitem{sdsharma} S. D. Sharma $et~al.$, NJP Quantum Information {\bf 1}, 15001 (2015).
\bibitem{qiu} H. Qiu $et~al.$, Phys. Rev. Lett. 132, 186601 (2024).
}

   
 



\end{thebibliography}
\end{document}